\documentclass{ws-mpla}
\def\Journal#1#2#3#4{{#1} {\bf #2}, #3 (#4)}

\def\RNC{\em Rivista Nuovo Cimento}

\def\NIMA{{\em Nucl. Instrum. Methods} A}

\def\PLB{{\em Phys. Lett.}  B}
\def\PRL{\em Phys. Rev. Lett.}
\def\PRD{{\em Phys. Rev.} D}

\def\GaC{\em Gravitation and Cosmology}
\def\GaCS{{\em Gravitation and Cosmology} Supplement}

\def\JETPL{\em JETP Lett.}
\def\PAN{\em Phys.Atom.Nucl.}
\def\CQG{\em Class. Quantum Grav.}
\def\APJ{\em Astrophys. J.}
\def\SCI{\em Science}
\def\MPLA{{\em Mod. Phys. Lett.}  A}
\def\IJTP{\em Int. J. Theor. Phys.}
\def\IJMPA{{\em Int. J. Mod. Phys.}  A}
\def\IJMPD{{\em Int. J. Mod. Phys.}  D}
\def\NJP{\em New J. of Phys.}
\def\ARAA{\em Ann. Rev. Astron. Astrophys.}
\def\AIPCP{\em AIP Conf. Proc.}
\def\JHEP{\em JHEP}
\def\JCAP{\em JCAP}
\def\EPHJ{\em Eur.Phys.J}
\def\JPCS{{\em J. Phys.:} Conf. Ser.}
\def\BWP{\em Bled Workshops in Physics}

\def\({\left(}
\def\){\right)}

\def\beq{\begin{equation}}
\def\eeq{\end{equation}}
\def\bea{\begin{eqnarray}}
\def\eea{\end{eqnarray}}

\begin{document}

\markboth{M.Yu.KHLOPOV}
{INTRODUCTION}

\catchline{}{}{}{}{}

\title{INTRODUCTION TO THE SPECIAL ISSUE\\ "INDIRECT DARK MATTER SEARCHES".
}

\author{MAXIM YU. KHLOPOV}

\address{National Research Nuclear University "MEPHI" (Moscow Engineering Physics Institute) and \\
    Centre for Cosmoparticle Physics "Cosmion" 115409 Moscow, Russia \\
APC laboratory 10, rue Alice Domon et L\'eonie Duquet \\75205
Paris Cedex 13, France\\
khlopov@apc.univ-paris7.fr}

\maketitle

\pub{Received (Day Month Year)}{Revised (Day Month Year)}

\begin{abstract}
The nature of cosmological dark matter finds its explanation in physics beyond the Standard model of elementary particles. The landscape of dark matter candidates contains a wide variety of species, either elusive or hardly detectable in direct experimental searches. Even in case, when such searches are possible the interpretation of their results implies additional sources of information, which provide indirect effects of dark matter. Some nontrivial probes for the nature of the dark matter are presented in the present issue.

\end{abstract}

\ccode{PACS Nos.: include PACS Nos.}

Cosmological dark matter is the important basic element of our current understanding of the structure and evolution of the Universe. It cannot be explained by the known forms of matter and implies physics beyond the Standard model of elementary particles for its description.
The list of dark matter candidates, predicted by the extensions of the Standard model, is so wide that can be hardly covered in an issue like that. Still rather wide range of possibilities is considered here: primordial black holes, axions and axion-like particles, decaying, annihilating and composite dark matter. Various candidates imply different ways to probe their existence and there is no need to repeat in this Introduction the content of the contributions to the present issue. Instead we'd like here to give a brief review of the general frame of the fundamental relationship of macro- and micro worlds, in which indirect effects of dark matter play important role.

The problem of dark matter, corresponding to
$\sim 25\%$ of the total cosmological density, involves two aspects,
in which physics beyond the standard model of elementary particles is involved.
The dark matter species (see e.g.\cite{book,newBook,DMRev,DDMRev}
for review and reference) should
be stable, saturate the measured dark matter density and decouple
from plasma and radiation at least before the beginning of matter
dominated stage. The latter condition is needed to make compatible formation of the Large scale structure of the Universe with the observed level of CMB fluctuations.
Therefore formation of the Large Scale Structure of the Universe from
small initial density fluctuations is one of the most
important reasons for the {\it nonbaryonic} nature of the dark matter
that is decoupled from matter and radiation and provides the effective growth of these fluctuations before recombination. On the other hand, the initial density fluctuations, coming from the very early Universe are also originated from physics beyond the Standard model. It puts together these two aspects of the dark matter problem.

Moreover, the stage, the expanding Universe, on which dark matter acts, involves several other basic elements, whose physical nature goes beyond the known laws of physics. The modern cosmological picture is based on inflational models, explaining the global properties of the Universe, with baryosynthesis, explaining the lack of antimatter in it, and dark energy, dominating in the modern Universe and causing its currently observed accelerated expansion. All these phenomena - inflation, baryosynthesis and dark energy - also find their nature and mechanisms in the extensions of the Standard model.
It puts the problem of dark matter and its physical nature in the multi-parameter space of new physics and one must find a way, how we can make overdetermined the system of equations relative to the unknown parameters, so that the true picture of the cosmological structure and evolution as well the physical laws governing them can be revealed. The approach to the solution of this multi-dimensional problem implies all the possible sources of information about possible signatures of new physics and combination of indirect evidences is crucial for probe its laws and features.

Let's give briefly following \cite{DMRev,DDMRev} some samples of dark candidates. Even remaining in our 1+3 dimensional space-time and not involving prediction of theories with infinite extra dimensions we find a wide variety in the description of the physical nature of dark matter, implying different specific means for its probe.

Most of the known particles are unstable. For a particle with the
mass $m$ the particle physics time scale is $t \sim 1/m$
\footnote{Here and further, if it isn't specified otherwise we use the units $\hbar=c=k=1$}, so in
particle world we refer to particles with lifetime $\tau \gg 1/m$
as to metastable. To be a candidate for dark matter the particle should survive in the Universe to the present time, so it should be either absolutely stable, or have a lifetime, exceeding the age of the Universe.
Such a long (or even infinite) lifetime should find reason in the existence of an (approximate)
symmetry. From this viewpoint, the dark matter cosmology is sensitive to the most
fundamental properties of microworld, to the conservation laws
reflecting strict or nearly strict symmetries of particle theory.

\subsubsection*{Stable relics. Freezing out. Charge symmetric case}
 \label{WIMPs}
The simplest form of dark matter candidates is the gas of new
stable neutral massive particles, originated from early Universe. For
particles with the mass $m$, at high temperature $T>m$ the
equilibrium condition, $$n \cdot \sigma v \cdot t > 1$$ is valid, if
their annihilation cross section $\sigma > 1/(m m_{Pl})$ is
sufficiently large to establish the equilibrium. At $T<m$ such
particles go out of equilibrium and their relative concentration
freezes out. This
is the main idea of calculation of primordial abundance for
Weakly Interacting Massive Particles (WIMPs, see e.g. Refs.
\cite{book,newBook,DMRev} for details), which provide realization of the Cold Dark Matter (CDM) scenario.

The process of WIMP annihilation to ordinary particles, considered in $t$-channel,
determines their scattering cross section on ordinary particles and thus
relates the primordial abundance of WIMPs to their scattering rate in the
ordinary matter. Forming nonluminous massive halo of our Galaxy, WIMPs can penetrate
the terrestrial matter and scatter on nuclei in underground detectors. The strategy of
direct WIMP searches implies detection of recoil nuclei from this scattering.

The process inverse to annihilation of WIMPs corresponds to their production in collisions
of ordinary particles. It should lead to effects of missing mass and energy-momentum,
being the challenge for experimental search for production of dark matter candidates at accelerators,
e.g. at the LHC.

\subsubsection*{Stable relics. Decoupling}

More weakly interacting and/or more light species decouple from plasma
and radiation being relativistic
at $T \gg m$, when $$n \cdot \sigma v \cdot t \sim 1,$$
i.e. at $$T_{dec} \sim (\sigma m_{Pl})^{-1} \gg m.$$ After decoupling these species retain
their equilibrium distribution until they become non-relativistic at $T < m$.
Conservation of partial entropy in the cosmological expansion links the modern abundance
of these species to number density of relic photons with the account for the increase of
the photon number density due to the contribution of heavier ordinary particles, which were
in equilibrium in the period of decoupling.
\subsubsection*{Stable relics. SuperWIMPs}
The maximal
temperature, which is reached in inflationary Universe, is the
reheating temperature, $T_{r}$, after inflation. So, the very
weakly interacting particles with the annihilation cross section
$$\sigma < 1/(T_{r} m_{Pl}),$$ as well as very heavy particles with
the mass $$m \gg T_{r}$$ can not be in thermal equilibrium, and the
detailed mechanism of their production should be considered to
calculate their primordial abundance.

In particular, thermal production of gravitino in very early Universe is proportional to the reheating temperature $T_{r}$, what puts upper limit on this temperature from constraints on primordial gravitino abundance\cite{khlopovlinde,khlopovlinde2,khlopovlinde3,khlopov3,khlopov31,Karsten,Kawasaki}.

\subsubsection*{Self interacting dark matter}\label{mirror}
Extensive hidden sector of particle theory can provide the existence of new interactions, which only dark sector particles possess. Historically one of the first examples of such self-interacting dark matter was presented by the model of mirror matter. Mirror particles, first proposed by T. D. Lee and C. N. Yang in Ref. \cite{LeeYang} to restore equivalence of left- and right-handed co-ordinate systems in the presence of P- and C- violation in weak interactions, should be strictly symmetric by their properties to their ordinary twins. After discovery of CP-violation it was shown by I. Yu. Kobzarev, L. B. Okun and I. Ya. Pomeranchuk in Ref. \cite{KOP} that mirror partners cannot be associated with antiparticles and should represent a new set of symmetric partners for ordinary quarks and leptons with their own strong, electromagnetic and weak mirror interactions. It means that there should exist mirror quarks, bound in mirror nucleons by mirror QCD forces and mirror atoms, in which mirror nuclei are bound with mirror electrons by mirror electromagnetic interaction \cite{ZKrev,FootVolkas}. If gravity is the only common interaction for ordinary and mirror particles, mirror matter can be present in the Universe in the form of elusive mirror objects, having symmetric properties with ordinary astronomical objects (gas, plasma, stars, planets...), but causing only gravitational effects on the ordinary matter \cite{Blin1,Blin2}.

Even in the absence of any other common interaction except for gravity, the observational data on primordial helium abundance and upper limits on the local dark matter seem to exclude mirror matter, evolving in the Universe in a fully symmetric way in parallel with the ordinary baryonic matter\cite{Carlson,FootVolkasBBN}. The symmetry in cosmological evolution of mirror matter can be broken either by initial conditions\cite{zurabCV,zurab}, or by breaking mirror symmetry in the sets of particles and their interactions as it takes place in the shadow world\cite{shadow,shadow2}, arising in the heterotic string model. We refer to Refs.
\cite{newBook,okun,Paolo} for current review of mirror matter and its cosmology.

\subsubsection*{Subdominant dark matter}
If charge symmetric stable particles (and their antiparticles) represent
only subdominant fraction of the cosmological dark matter, more detailed analysis
of their distribution in space, of their condensation in galaxies,
of their capture by stars, Sun and Earth, as well as effects of
their interaction with matter and of their annihilation provides
more sensitive probes for their existence. It was first revealed in\cite{ZKKC} and then proved in\cite{DKKM} for subdominant particles that the contribution of dark matter annihilation in the flux of cosmic ray positrons provides a mighty indirect probe for dark matter.

\subsubsection*{Decaying dark matter}
Decaying particles with lifetime $\tau$, exceeding the age of the
Universe, $t_{U}$, $\tau > t_{U}$, can be treated as stable. However, even small effect of their decay
can lead to significant contribution to cosmic rays and gamma background\cite{ddm}.
Leptonic decays of dark matter are considered as possible explanation of
the cosmic positron excess, measured in the range of hundreds GeV by PAMELA\cite{pamela}, FERMI/LAT\cite{lat} and AMS02\cite{ams21,ams22}.

Primordial unstable particles with the lifetime, less than the age
of the Universe, $\tau < t_{U}$, can not survive to the present
time. But, if their lifetime is sufficiently large, their existence in
early Universe can lead to observable effects\cite{khlopov7}.

Weakly interacting particles, decaying to invisible modes, can influence Large Scale Structure formation.
Such decays prevent formation of the structure, if they take place before the structure is formed.
Invisible products of decays after the structure is formed should contribute in the cosmological dark energy.
The Unstable Dark matter scenarios\cite{Sakharov1,UDM,UDM1,UDM2,UDM3,berezhiani4,berezhiani5,TSK,GSV} implied weakly interacting particles that form the structure on the matter dominated stage and then decay to invisible modes after the structure is formed.

Cosmological
flux of decay products contributing into the cosmic and gamma ray
backgrounds represents the direct trace of unstable particles\cite{khlopov7,sedelnikov}. If
the decay products do not survive to the present time their
interaction with matter and radiation can cause indirect trace in
the light element abundance\cite{khlopovlinde3,khlopov3,khlopov31,DES} or in the fluctuations of thermal
radiation\cite{UDM4}.
\subsubsection*{Charge asymmetry of dark matter}
The fact that particles are not absolutely stable means that the corresponding charge is not strictly conserved and generation of particle charge asymmetry is possible, as it is assumed for ordinary baryonic matter. At sufficiently large particle annihilation cross section excessive particles (antiparticles) can dominate in the relic density, leaving exponentially small admixture of their antiparticles (particles) in the same way as primordial excessive baryons dominate over antibaryons in baryon asymmetric Universe. In this case {\it Asymmetric dark matter} doesn't lead to significant effect of particle annihilation in the modern Universe and can be searched for either directly in underground detectors or indirectly by effects of decay or condensation and structural transformations of e.g. neutron stars (see Ref. \cite{adm} for recent review and references).
\subsubsection*{Charged stable relics. Dark atoms}
New particles with electric charge and/or strong interaction can
form anomalous atoms and contain in the ordinary matter as anomalous
isotopes. For example, if the lightest quark of 4th generation is
stable, it can form stable charged hadrons, serving as nuclei of
anomalous atoms of e.g. anomalous helium
\cite{BKSR1,BKS,BKSR,FKS,I,BKSR4}. Therefore, stringent upper limits on anomalous isotopes, especially, on anomalous hydrogen put severe constraints on the existence of new stable charged particles. However, stable doubly charged particles can not only exist, but even dominate in the cosmological dark matter, being effectively hidden in neutral "dark atoms"\cite{DMRev,DDMRev,DADM}.
\subsubsection*{Phase transitions}
Parameters of new stable and metastable particles are also
determined by a pattern of particle symmetry breaking. This pattern
is reflected in a succession of phase transitions in the early
Universe. First order phase transitions proceed through bubble
nucleation, which can result in black hole formation (see e.g.
Ref. \cite{book2,PBHrev} for review and references). Phase
transitions of the second order can lead to formation of topological
defects, such as walls, string or monopoles. Structure of cosmological defects can be
changed in succession of phase transitions. More complicated forms
like walls-surrounded-by-strings can appear. Such structures can be
unstable, but their existence can leave a trace in nonhomogeneous
distribution of dark matter.

\subsubsection*{Axions and axion-like particles}
Succession of phase transitions, correponding to the spontaneous and then manifest breaking of U(1) symmetry, results in the generation in the Universe of a pseudo-Nambu--Goldstone (PNG) field (see Refs. \cite{DMRev,book2,PBHrev} for review and references). The coherent oscillations of this field represent a specific type
of CDM in spite of a very small mass of PNG particles $m_a=\Lambda^2/f$, where $f \gg \Lambda$, since these particles are created in Bose-Einstein condensate in the ground state, i.e. they are initially created as nonrelativistic in the very early Universe.
This feature, typical for invisible axion models can be the general feature for all the axion-like PNG particles.

At high temperatures the pattern of successive spontaneous and manifest breaking of global U(1) symmetry implies the
succession of second order phase transitions. In the first
transition at $T \sim f$, continuous degeneracy of vacua leads, at scales
exceeding the correlation length, to the formation of topological
defects in the form of a string network; in the second phase
transition at $T \sim \Lambda \ll f$, continuous transitions in space between degenerated
vacua form surfaces: domain walls surrounded by strings. This last
structure is unstable, but, as was shown in the example of the
invisible axion \cite{Sakharov2,kss,kss2}, it is reflected in the
large scale inhomogeneity of distribution of energy density of
coherent PNG (axion) field oscillations. This energy density is
proportional to the initial value of phase, which acquires dynamical
meaning of amplitude of axion field. The value of phase changes by $2 \pi$ around string. This strong
nonhomogeneity of phase leads to corresponding nonhomogeneity of
energy density of coherent PNG (axion) field oscillations, which is correlated at large distances as a replica of the original structure of unstable topological defects - walls-surrounded-by-strings.
\subsubsection*{Primordial Black Holes}
 \label{pbhs}
 Any object of mass
$M$ can become a black hole, being put within its gravitational
radius $r_g=2 G M/c^2.$ At present time black holes can be
created only by a gravitational collapse of compact objects with
mass more than about three Solar mass \cite{1,ZNRA}. It can be a
natural end of massive stars or can result from evolution of dense
stellar clusters. However in the early Universe there were no limits
on the mass of a black hole. Ya.B. Zeldovich and I.D. Novikov (see Ref. \cite{ZN})
noticed that if cosmological expansion stops in some region, black
hole can be formed in this region within the cosmological horizon.
It corresponds to strong deviation from general expansion and
reflects strong inhomogeneity in the early Universe.

For long time scenarios with Primordial Black Holes (PBH) belonged
dominantly to cosmological {\it anti-Utopias}, to "fantasies", which
provided restrictions on physics of very early Universe from
contradiction of their predictions with observational data. Even
this "negative" type of information makes PBHs an important
theoretical tool.

By construction astrophysical constraint excludes effect, predicted
to be larger, than the observed one. At the edge such constraint converts
into an alternative mechanism for the observed phenomenon. At some
fixed values of parameters, PBH spectrum can play a positive role
and shed new light on the old astrophysical problems.

The common sense is to think that PBHs should have small sub-stellar
mass. Formation of PBHs within cosmological horizon, which was very
small in very early Universe, seem to argue for this viewpoint.
However, phase transitions on inflationary stage can provide spikes
in spectrum of fluctuations at any scale, or provide formation of
closed massive domain walls of any size.

Being far from complete, the above listing gives the flavor of the variety of possible features of the new physics in the cosmological dark matter.
The new physics follows from the necessity to extend the Standard
model. The white spots in the representations of symmetry groups,
considered in the extensions of the Standard model, correspond to
new unknown particles. The extension of the symmetry of gauge
group puts into consideration new gauge fields, mediating new
interactions. Global symmetry breaking results in the existence of
Goldstone boson fields.

In particle physics direct experimental probes for the predictions
of particle theory are most attractive. The predictions of new
charged particles, such as supersymmetric particles or quarks and
leptons of new generation, are accessible to experimental search
at accelerators, if their masses are in
100GeV-1TeV range. However, the predictions related to higher
energy scale need non-accelerator or indirect means for their
test and the observational cosmology offers strong indirect evidences favoring the
existence of processes, determined by new physics.

Cosmoparticle physics \cite{book,newBook}, studying the
physical, astrophysical and cosmological impact of new laws of
Nature, explores the new forms of matter and their physical
properties in the combination of their indirect effects. Physics of dark matter in all its aspects
plays important role in this process. It makes indirect dark matter searches of special interest and the present issue can be considered as the first step towards their extensive and systematic treatment. Our future planned special issue of Modern Physics Letters A "Dark Matter Candidates" and Book Review "Indirect Effects of Dark Matter Physics" will continue exploration of this exciting field of scientific research.

\section*{Acknowledgments}
I express my gratitude to E. Nash and E. H. Chionh for kind proposal to be the Guest Editor of this issue and to all the authors of the contribution to it. The work on initial cosmological conditions was supported by the Ministry of Education and Science of Russian Federation, project 3.472.2014/K and the work on the forms of dark matter was supported by grant RFBR 14-22-03048.

\end{document}